# The 6th Law of Stupidity: A Biophysical Interpretation of Carlo Cipolla's Stupidity Laws

Ilaria Perissi and Ugo Bardi[1]
Dipartimento di Chimica – Università di Firenze.
Polo Scientifico di Sesto Fiorentino, via della Lastruccia 3
50019 Sesto Fiorentino (Fi) - Italy

## Abstract

Carlo Cipolla's "stupidity quadrant" and his five laws of stupidity were proposed for the first time in 1976 [1]. Exposed in a humorous mood by the author, these concepts nevertheless describe very serious features of the interactions among human beings. Here, we propose a new interpretation of Cipolla's ideas in a biophysical framework, using the well-known "predator-prey," Lotka-Volterra model. We find that there is indeed a correspondence between Cipolla's approach – based on economics – and biophysical economics. Based on this examination, we propose a "6th law of stupidity," additional to the five proposed by Cipolla. The law states that "humans are the stupidest species in the ecosystem."

1. Introduction

In 1976, the economist and historian Carlo M. Cipolla (1922-2000) wrote an essay titled "*The Basic Laws of Human Stupidity*." Initially, it was only a pamphlet circulated among friends [1], but later it was published as a book [2]. Written in a tongue-in-cheek style, Cipolla's text analyzed human behavior using a simple semi-quantitative model in the form of two individuals ("agents") interacting with each other in performing an economic transaction.

Cipolla reasoned in terms of the payoff of each transaction, arranging the possible outcomes as a quadrant divided into four subsectors. One of the two agents may gain something at the expense of the other, but it may also happen that both profit from the exchange. The worst possible situation is the one in which both lose something. The kind of agents who cause someone else's loss while damaging also themselves in the process were labeled by Cipolla as "stupid people."

From there, Cipolla went on defining the five "laws of stupidity" as 1) Always and inevitably everyone underestimates the number of stupid individuals in circulation. 2) The probability that a certain person will be stupid is independent of any other characteristic of that person. 3) A stupid person is a person who causes losses to another person or to a group of persons while himself deriving no gain and even possibly incurring losses, 4) Non-stupid people always underestimate the damaging power of stupid individuals, and 5) A stupid person is the most dangerous type of person.

---

[1] Corresponding author: ugo.bardi@unifi.it

Today, Carlo Cipolla may well be better known for his quadrant and the five laws, that he probably thought of as a joke, than for his academic papers. One of the reasons for this popularity is that these ideas ring true: they make sense according to our everyday experience. Indeed, Cipolla's ideas have been examined, discussed, and modeled in various ways for instance in terms of game theory [3] and of agent-based modeling [4].

Here, we wish to take a fresh look at Cipolla's theory using a biophysical approach. That is, we will frame Cipolla's quadrant in terms of a complex system similar to biological ones. We'll use the model known as the "Lotka-Volterra" (LV) one, also known as the "predator-prey" or "Foxes and Rabbits" model [5], [6]. Our examination leads us to propose a "6$^{th}$ law of stupidity" that applies to the whole ecosystem and that has that "Humans are the stupidest species on Earth."

2. The Model

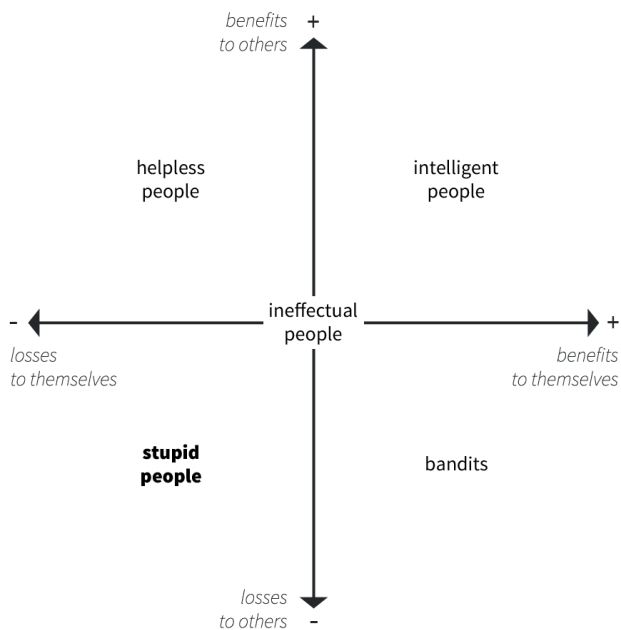

Figure 1. *"Cipolla's Quadrant" (*[7]*)*

Cipolla's approach is illustrated by the quadrant shown in fig. 1. The orthogonal axes refer to two agents participating in an economic transaction: positive values indicate a gain, negative values a loss. Typically, the x-axis is referred to the observer (you), while the y-axis to the second agent ("the other person")

Starting with the top left, we have a situation in which "helpless people" damage themselves while providing an advantage to the other person. In the original paper [1], Cipolla made the example of selling an old and valueless pig to another person for a high price. The top right quadrant describes a deal that benefits both participants: you sell a good pig at an honest price to the other person. The bottom right quadrant is that of the bandits: you steal a pig or pay it much less than it is worth. Finally, the bottom left quadrant is the one dealing with stupidity, a phenomenon that occurs when someone's action results in damage to both parties involved. Stupidity does not rule out temporary gains of the stupid agent or even of both agents, but just that the actions of one of the agents will eventually bring ruin to both. In Cipolla's original paper [1], stupidity was not illustrated with images of people selling or buying pigs, but we might build up a narrative involving pigs imagining someone who kills the owner of a pig in order to steal it, and then kills himself and the pig in a road crash while taking home his ill-gotten gain.

The model can also be seen in wider terms, that is not just describing monetary transactions in the economy. No economic process could take place without some energy being available. So, we may see interactions among agents as involving an exchange of energy. In this case, the agents are not necessarily human beings but may be individuals or populations of biological species involved in interactions with other species. This view is part of the general concept of "biophysics" that deals with dynamic complex systems. More exactly, the economic process is a form of dissipation of energy potentials [8]. In this sense, it is not different from most dynamic processes ongoing in the universe.

Unsurprisingly, the similarities between the economic process and the biological process appear in Cipolla's quadrant. We can re-examine the four sectors in this sense. Clockwise from the upper left quadrant, we have that:

1. *Helpless people* → Predator and Prey: here, the predator gains metabolic energy from the prey and the prey gains nothing. It can also be seen in terms of a parasite/host relationship.
2. *Intelligent people* → Symbiosis. In this case, we deal with "symbionts." It is a partnership of two species that may involve an unbalanced exchange of energy but that, overall, benefits both partners.
3. *Bandits* → Prey and Predator. This is equivalent to case 1, apart from exchanging the roles of predator and prey.
4. *Stupid people* → Those parasites which kill the host or predators that exterminate the whole prey population. Note that the parasites usually obtain a short-term gain in this way, but that is nullified by their eventual ruin for lack of food.

Having established these similarities, we can interpret Cipolla's quadrant using one of the simplest population models known in biology, the one developed by Alfred Lotka [5] and Vito Volterra [6] in the 1920s.

The Lotka-Volterra (LV) model describes the interaction of two populations, predators and prey, often referred to as "rabbits" and "foxes." The two species form a two-stage trophic chain where metabolic energy is dissipated in steps: The first level (rabbits) acquires energy from a source supposed to be unlimited (grass). This energy is transferred to the second level (foxes) by the process of predation. Finally, it is dissipated to the environment as low-temperature heat.

In its simplest form, the LV model is described by two coupled differential equations.

*$dL_1/dt = aL_1 - bL_1L_2$*

*$dL_2/dt = \eta bL_1L_2 - cL_2$*

$L_1$ represents the population of the 1st trophic level (the prey, or rabbits). $L_2$ is the population of the 2nd level (the predators, or foxes). *a, b,* and *c* are constant coefficients, *η* is an efficiency parameter that ranges from 0 to 1. In the standard version of the model, all these coefficients are positive.

In the "competitive" version of the model, a further parameter is introduced: the carrying capacity of the system. This parameter sets a limit to the maximum size that the two populations can reach. The equations become:

*$dL_1/dt = aL_1(1-a/N) - bL_1L_2$*

*$dL_2/dt = \eta bL_1L_2 - cL_2$*

With *N* measuring the maximum level of the $L_1$ population. Another limiting factor may be added to the 2nd equation, but here we will assume that the $L_2$ stock is limited mainly by the size of the $L_1$ population.

A characteristic of the LV system is the presence of "feedbacks" that govern the energy transfers: the energy transfer rate is proportional to the size of the stocks. Because of this, we can say that this system is *autocatalytic*. The more a stock grows, the more it draws energy from another stock or from the environment, a behavior typical of biological populations and also of economic entities such as companies and corporations.

Of course, neither populations nor corporations can keep growing forever because they exploit finite resources (the concept of "spaceship Earth" [9]). The LV model takes into account the limits to growth of the system by having the growth rate proportional to the product of the predator and prey stocks. If the predators overexploit the prey, growth will slow down and will eventually turn into decline. The result described by the LV model is that the two populations tend to go through cycles of rapid growth and sudden collapses. This condition, termed "homeorhesis" [10], [11], means that the populations follow a trajectory that never strays away too much from an average called the "attractor" of the system. In some variants of the model (the so-called "competitive" version), the attractor is reached, and the unperturbed system maintains its parameters constant in a condition called "homeostasis."

The Lotka-Volterra model is normally considered too simple to be able to provide a quantitative description of real ecosystems. However, it has been shown to describe several real-world cases, not just biological ones. It has been applied to oil extraction [12], the 19th-century whaling industry [13], and the production cycle of modern fisheries [14]. It can also describe the diffusion of memes in the World Wide Web [15]. It is a versatile model: playing with the signs of the coefficients it can describe widely different behaviors such as the attraction of Giulietta and Romeo in Shakespeare's play [16] [17]

In a biologic system, the two levels of the trophic chain exchange metabolic energy under the effect of energy potentials. In Cipolla's model, the agents exchange money or goods under what we may call a "financial potential." So, we can take the two LV levels as representing the two agents of the Cipolla model, with the levels of the stocks representing their wealth.

Now, we need to describe how the LV model can be tuned to reproduce the characteristics of the interactions described by Cipolla. We will examine each quadrant separately.

*1st and 3rd Quadrants: bandits and their helpless victims*

The two diagonally opposite quadrants of Cipolla's model, those labeled "Helpless," and "Bandits," are the same thing. In both cases, we have a prey/predator interaction, with the bandit as the predator and the helpless agent as the prey. The only difference is the viewpoint of an observer. In biology, "bandit species" are rare, but they do exist, and the result is oscillations in the prey and predator populations. A good example is that of the budworm infestations of spruce forests described by Holling [18].

In economics, the bandit/victim system can be described by the standard LV model with $L_1$ corresponding to the victim and $L_2$ to the bandit. We could also say that the bandits and the victims are in a seller/buyer relationship, but either the seller or the buyer are forcing the counterpart into a bad deal. The figure below shows how the system evolves in time according to the LV equations, solved iteratively. In the figure, the y-scale represents the level of the stocks and it can be seen as a measure of the monetary wealth of the agents: bandits and victims.

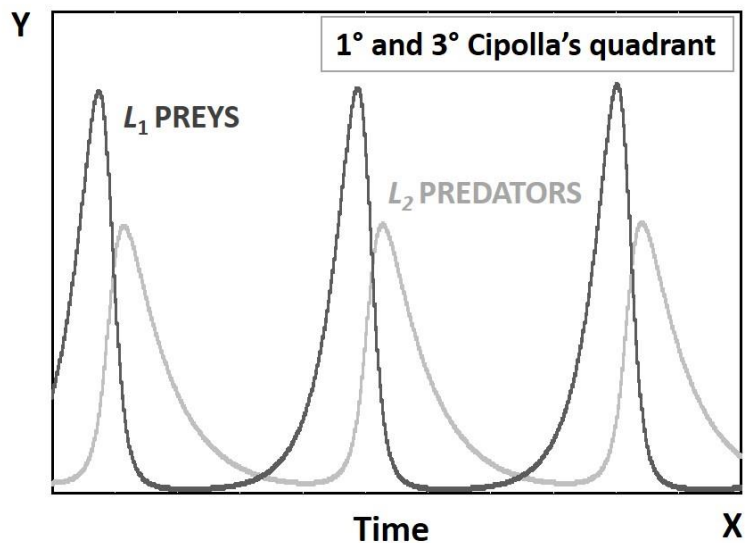

*Figure 2. Bandits and their helpless victims according to the LV model.*

A qualitative interpretation of these results is that the bandits build-up their wealth by depredating (or swindling) their helpless victims, who become poorer as a consequence. At some point, the victims are so impoverished that the bandits do not make any more a profit by attacking or dealing with them. That forces the bandits to reduce their activity, maybe moving to other sources of revenue. At this point, the victims have a chance to rebuild their assets and become again interesting targets for robbery.

More in detail, the model tells us that the bandits operate at a certain cost. Just like predators need metabolic energy to outrun their prey, bandits need weapons, information, transportation, and more. From the equations of the model, we can say that banditry will be profitable if $dL_2/dt > 0$. This condition implies that $\eta b L_1 / c > 1$. The term $\eta b L_1 / c$ is the return on the investment in robbery, ROI. In a biophysical system, it is called EROI or EROEI (energy return on energy invested) [19].

According to Cipolla, a "perfect" bandit is one whose gains in the robbery equals exactly the loss of the victim. In the LV model, such a bandit would be described for an efficiency $\eta=1$, impossible to obtain in the real world because of the limits set by thermodynamics. Bandits will nevertheless try to maximize their ROI by being as efficient as possible ($\eta$ close to 1). They may also try to decrease the value of the *c* coefficient, decreasing the cost of banditry. Another possible strategy for them is to increase the value of the *b* coefficient, increasing the intensity of their robbing activity. Symmetrically, the victims may attempt to react to bandits by rebuilding their wealth as fast as possible, that is increasing the *a* coefficient. These attempts will not eliminate the oscillations of the systems, they may only affect their frequency that varies as $\omega=\sqrt{ac}$.

The problem with the bandits' approach is that the system can reach a stable state only if two conditions occur at the same time, $dL_1/dt= 0$ and $dL_2/dt= 0$, which does not normally happen, except for special values of the coefficients. Apart from this special case, the system will keep oscillating without ever reaching stability. That will make it impossible for the bandits to stably accumulate wealth. It is because they are implicitly assumed to optimize their short-term gains, so they will not make any attempt to plan for a long-term future that would avoid ruining their victims. As a consequence, the ROI of banditry is destined to decline in each cycle, since it is proportional to the wealth of the victims. The model tells us that, as it is obvious from empirical evidence, it is not a good idea to rob the same person twice in a row.

*2nd Quadrant: Intelligent agents*

This quadrant in Cipolla's model represents a condition in which both actors involved in a transaction gain something. In a biological system, this quadrant describes the condition known as *symbiosis* – two species that form an alliance that benefits both in terms of energy exchange. Symbiotic systems are well-known in the ecosystem, a classic example is the gut microbiota of human beings that receives food from the human organism and helps it to digest it. This system is described by Lynn Margulis using the term "holobiont" [20]. Holobionts are normally understood to originate as the result of a gradual process of adaptation. It may well be that a competitive predator/prey interaction evolves into a symbiosis after many cycles of destructive interactions where the two species gradually "learn" how to respect each other and the physical limits of the system.

The Lotka-Volterra model cannot directly describe evolution, but it does depend on the limits of the system. We may see the two-species trophic chain as a control system that regulates the population of both rabbits and foxes. Note that all control systems must have a memory, otherwise they would not "know" what level the system should tend to reach [21]. In the simplest form of the LV model, the memory is not a separate stock, as it is in most human-made control devices. It is, rather, stored in the two population stocks. The size of the rabbit stock depends on the size of the fox stock and vice versa and the two stocks regulate each other. The regulation is not perfect but, at least, the two stocks never exceed certain limits.

In the competitive version of the LV model, a further element is added: the maximum size of at least one of the stocks (the "carrying capacity" of the system). The effect of this term is expressed by multiplying the $L_1$ stock size in the first equation of the model by a factor ($1-L_1/N$), with N the carrying capacity. This term affects the dependency of the population growth on the population size making it not anymore linear, but diminishing as $L_1$ approaches N. The result is that the system converges to a well-defined attractor in the form of two specific values of the stocks.

N is a "memory element" that tells the system how to control itself. But how does the system know that? Evidently, in real ecosystems there are physical signals that regulate the growth of both stocks. For instance, rabbits face a dearth of grass that prevents them from growing exponentially even in the absence of foxes. But that is not the only way to transfer this information to the system and we may imagine intelligent agents who can plan and try to optimize the system even before growth is reined in by physical limitations. In the evolutionary game, these agents would be favored in the long run because they could avoid the crash that follows excessive growth that might bring competitors to extinction. These considerations cannot be proven, but they provide some justification to the idea of using the competitive Lotka-Volterra model to describe the Cipolla quadrant dealing with intelligent people.

In an economic system, the considerations about biological systems may be seen as due to intelligent bandits who can understand that over-exploiting their victims is not a good thing for them. Limiting the exploitation activity could also be the result, for instance, of progressive taxation or of the government introducing quotas or other balancing factors in the exploitation of natural resources – again actions performed by intelligent agents. Once this assumption is inserted in the equations of the model in the form of a finite carrying capacity, we see that the agents/populations tend to reach a condition of homeostasis in terms of stable wealth.

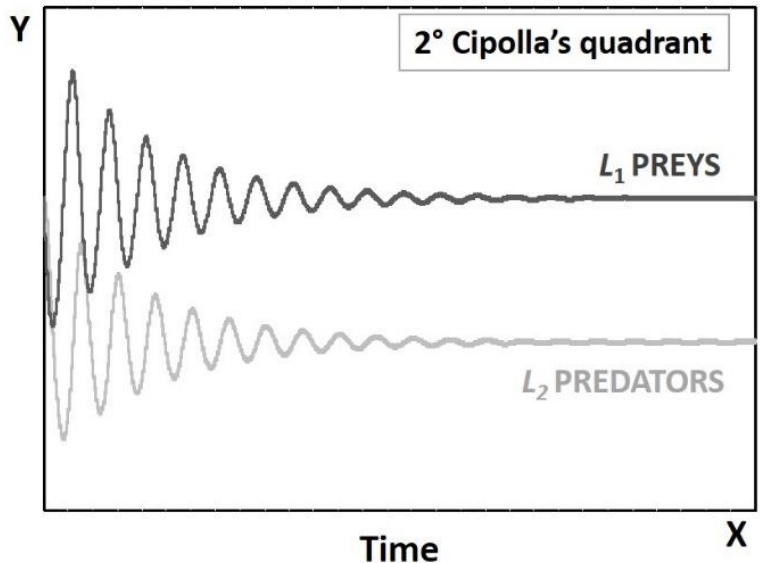

Figure 3. Evolution of the LV model for the 2nd quadrant of Cipolla's model. In this case, the preys are buyers while the sellers are the predators.

The initial oscillations reflect an initial phase of reciprocal adaptation. Then, an intelligent buyer and an intelligent seller will agree on a price that benefits both. In this system, there holds the same condition as for the bandits/helpless people (predators/prey) quadrants. That is the ROI/EROI of the transaction is $\eta b L_1/c$ which is equal to 1 in homeostatic conditions. In this case, the homeostasis condition leads to the requirement that $dL_1/dt = 0$. The consequence is that $a(1-N)-bL_2 = 0$. That is, $L_2 = a(1-a/N)/b$. Bandits will ensure their maximum wealth by making sure that this expression is maximized, which involves $a = N/2$, that is $L_2 = N/4b$. Since we can express b as a function of $L_1$ we can obtain a relation between $L_1$ and $L_2$ at homeostasis. The result is that $L_2 = \eta N L_1 / 4c$.

This result implies that to maximize their wealth, bandits should be as efficient as possible ($\eta$ close to 1). They should also reduce their costs (*c* as small as possible), and make sure that the carrying capacity of the system is as large as possible. Note that the formula implies that the wealth of the bandits is proportional to that of their victims. That agrees with the common wisdom that bandits should always rob rich people (something that Robin Hood had understood without the need for mathematical models). But the LV model also tells us that, as a victim, you should always prefer to be robbed by a rich bandit!

### 4th Quadrant: Stupid people

In a biological system, the equivalent of stupidity according to Cipolla's definition implies that the parasites/predators destroy their host/prey. In biology, there exists the concept of "Parasitoids," parasites that kill their host. But, normally, parasitoids are not stupid: killing the host is part of a strategy that involves moving to another host.

Nevertheless, some parasites/predators can operate against their own survival and destroy themselves by killing their host/prey. This kind of behavior is rare in the ecosystem, but it exists: it is well known that viruses and bacteria can kill their hosts. An example involving vertebrates is the story of Matthew Island in the Northern Pacific Ocean, where a small number of reindeers were introduced by humans in 1944 [22]. Over little more than two decades, the reindeer grew in numbers, destroyed their sources of food, lichen and grass, and went extinct as a result. You may argue that the reindeer were not "stupid," it was rather the humans who introduced them to the island who were stupid. But this is exactly the point we are making in this paper.

In terms of the Cipolla quadrant, since the LV model assumes a continuous function for the values of the stocks, they can never go to zero and the model cannot simulate extinction or complete extermination of the prey. This limitation can be circumvented in various ways. The model can be easily modified by forcing the value of one of the stocks to go to zero when it becomes smaller than a fixed value. But a better way to simulate the stupidity quadrant is to assume that *a=0*. It means that the prey does not reproduce or reproduces so slowly that its regrowth can be neglected in the period of interest.

The results of the model calculations for the case a=0 are shown in the figure.

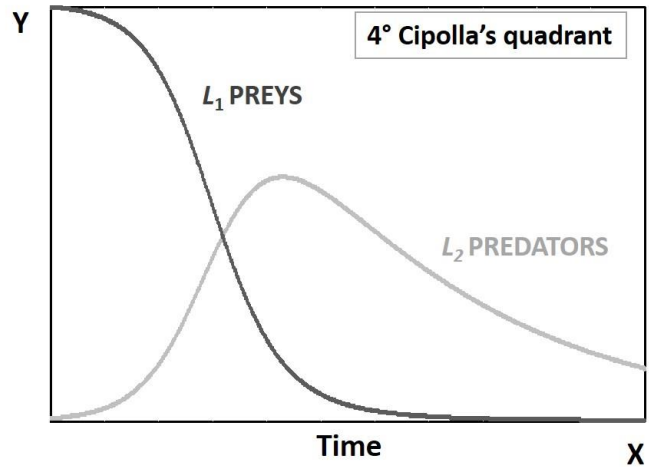

Figure 4. Results of the LV model for a=0

In the case of economic agents, just like in the case of bandits, stupid people do not optimize the system they exploit. But whereas the bandits can survive a crash in their revenues when their victims rebuild their wealth, stupid people ruthlessly destroy them, ruining themselves as well.

There are several examples in the history of economics: one is the case of the mining industry which is exploiting resources that will need at least hundreds of thousands of years to reform by geological process, if they ever will [23]. It is also the case of industries that exploit slowly reproducing biological resources. A modern example is that of whaling, as we demonstrated in previous papers [24], [12]. The same resource destruction also occurs for other cases of human fisheries [14]. Humans do not seem to need modern tools to destroy the resources they exploit, as shown by the extinction of Earth's megafauna [25], at least in part the result of human actions performed using tools not more sophisticated than stone-tipped spears.

Overall, the destruction of the resources that make people live seems to be much more common than in the natural ecosystem. This observation justifies the proposed "6[th] law of stupidity," additional to the five proposed by Carlo Cipolla that has that "Humans are the stupidest species in the ecosphere."

## 3. Conclusion

Cipolla's quadrant highlights several fundamental features of those systems that can be described as both "complex" and "autocatalytic," where the growth rate is proportional to the size of the stocks. These systems include living creatures, biomes, entire ecosystems, as well as human-created entities such as companies, organizations, and entire economic systems.

The analysis of Cipolla's quadrant, carried out using the Lotka-Volterra model shows the similarity of many phenomena driven by the dissipation of energy potentials: from life to commerce [26]. There are, indeed, some basic laws at work in these systems and when we use the term "law" for a physical system we mean that some factors are at work to keep it, if not perfectly regulated, at least within some boundaries.

Cipolla's quadrant tells us that these complex systems are all dominated by the same factors, but that these factors can operate in different ways. The simplest case is the predator/prey (bandit/victim) relationship, in which the predator seeks only maximum short-term profit. The result is periodical oscillations, homeorhesis. It is also possible to see the condition of "stupidity" where the actions of the actors in the exchanges lead to doom for everyone and everything. In ecosystems it is extinction, in economic systems, it is financial ruin.

The analysis also shows the possibility for these systems to adjust in such a way to attain the condition that Cipolla describes as "intelligent people" and that in ecosystems goes under the name of "symbiosis." As proposed by Lynn Margulis [20], symbiotic systems that go under the name of "holobionts" are the basic unit of the ecosystem. We may extend this definition to all kinds of autocatalytic complex systems, including those forming the human economy.

But if holobionts are an efficient unit of energy dissipation, why does stupidity exist? In particular, why is it so common in the economy as Cipolla correctly notes? Cipolla's description of stupid people is that

> *"..some are stupid and others are not, and that the difference is determined by nature and not by cultural forces or factors. One is stupid in the same way one is red-haired; one belongs to the stupid set as one belongs to a blood group. A stupid man is born a stupid man by an act of Providence."*

What Cipolla calls "an act of Providence" may be seen also as the result of the genetic setup of human beings. Indeed, humans are a relatively recent element of the ecosystem: modern humans are believed to have appeared only some 300,000 years ago, although other hominins practicing the same lifestyle may be as old as a few million years. Yet, this is a young age in comparison to that of most species currently existing in the ecosphere.

So, humankind's stupidity may be not much more than an effect of the relative immaturity of our species, which still has to learn how to live in harmony with the ecosystem. That explains what we called here "the 6$^{th}$ law of stupidity," stating that humans are the stupidest species on Earth. It is a condition that may lead the human species to extinction in a non-remote future. But it is also possible that, if humans survive, one day they will learn how to interact with the ecosystem of their planet without destroying it.

Acknowledgment. One of us (U.B.) would like to remember the figure of Carlo Maria Cipolla (1922-2000), whom he had a chance to meet in Berkeley in the 1980s. Cipolla was a brilliant and creative mind, but also a kind and open personality. His work, not just about stupidity, is still having an important impact on the way we see the world.

Conflict of interest. The authors have no conflict of interest related to this work. It was not specifically funded by any agency.